\RequirePackage{rotating}
\documentclass[format=acmsmall, review=false, screen=true]{acmart}
\usepackage{hvfloat}
\usepackage{acm-ec-18}
\usepackage{booktabs} 
\usepackage{balance}  
\usepackage{graphics} 
\usepackage{tabularx}
 \usepackage{amssymb}
 \usepackage{pifont}
\usepackage{tabulary}

\usepackage{rotating}
\usepackage{multirow}
\usepackage{diagbox}
\usepackage{hhline}
\newcommand{\cmark}{\ding{51}}%
\newcommand{\xmark}{\ding{55}}%
\newcolumntype{d}[1]{D{.}{.}{#1}}

\renewcommand{\paragraph}[1]{\hfill \break \textbf{#1.}}

\begin{document}
\title{Dancing Pigs or Externalities? Measuring the Rationality of Security Decisions}  
\author{Elissa M. Redmiles, Michelle L. Mazurek, and John P. Dickerson}
\affiliation{%
  \institution{University of Maryland}
 }
\email{[eredmiles,mmazurek,john]@cs.umd.edu}

\begin{abstract}
Accurately modeling human decision-making in security is critical to thinking about when, why, and how to recommend that users adopt certain secure behaviors. In this work, we conduct behavioral economics experiments to model the rationality of end-user security decision-making in a realistic online experimental system simulating a bank account. We ask participants to make a financially impactful security choice, in the face of transparent risks of account compromise and benefits offered by an optional security behavior (two-factor authentication). We measure the cost and utility of adopting the security behavior via measurements of time spent executing the behavior and estimates of the participant's wage. 
We find that more than 50\% of our participants made rational (e.g., utility optimal) decisions, and we find that participants are more likely to behave rationally in the face of higher risk. Additionally, we find that users' decisions can be modeled well as a function of past behavior (anchoring effects), knowledge of costs, and to a lesser extent, users' awareness of risks and context ($R^2$=0.61). We also find evidence of endowment effects, as seen in other areas of economic and psychological decision-science literature, in our digital-security setting. Finally, using our data, we show theoretically that a ``one-size-fits-all'' emphasis on security can lead to market losses, but that adoption by a subset of users with higher risks or lower costs can lead to market gains. 

\end{abstract}
\maketitle
%
\section{Introduction}
\label{sec:intro}
People's adoption, or rejection, of security behaviors can lead to system-wide consequences. While security requirements and system defaults such as automatic updates can remove many human decision-making fail-points for security-relevant systems, in many cases human decisions are still required to enable security. As such, security researchers have tried to understand and alter the decisions of the ``human in the loop''~\cite{cranor2008framework}.

Prior work has proposed two simplified theories of the human in the loop: a rational actor who chooses to ignore security behaviors because the costs always outweigh the potential losses, and an irrational actor who chooses ``dancing pigs over security every time'' because they neither understand nor care about security risks~\cite{Herley:2009:SLN:1719030.1719050}. While these simplified models of user behavior can help to provide high-level insights, our aim is to define a more realistic medium between these two extremes: \textit{a semi (or boundedly) rational security actor with predictable, but not always utility-optimal, behavior based on risks and costs}.

End-user decision-making in security has largely been examined hypothetically~\cite{Herley:2009:SLN:1719030.1719050, bohme2011security} or qualitatively~\cite{mathur2016they,vaniea2016tales,wash2015too,sheng2010falls,hong2012state,harbach2013sorry,sunshine2009crying,felt2014experimenting}. Here, we seek to provide an empirical, economic examination of the rationality of security behavior in a particular context. We define rationality as utility-optimality: that is, a decision is rational if the utility (gain) from the decision is greater than the costs of enacting the decision. Ultimately, we seek to understand: \textbf{How do costs ($C$), risks ($R$), and user tendencies and attributes ($U$) influence: (1) a security decision and (2) whether that decision is rational?} 

By answering this question and quantifying a model of rationality in security behavior, we can guide users toward utility-maximizing behavior (or lack of behavior) in order to optimize the use of their personal, behavioral compliance budgets and maximize market gains from security~\cite{Beautement:2009er,acquisti2005privacy}.

In this work, we construct an experimental system and conduct behavioral-economics experiments (e.g., games) to evaluate and model security decision-making. Such experiments allow for a high level of control, realism, and quantification beyond that usually achieved from traditional survey and lab-based user-studies~\cite{grossklags2007experimental}. Our experimental system operates like a bank account to which users have to regularly log in. In each game, users are assigned to a condition: in one of the endowment conditions participants are given an amount of money and are required to login once every 24 hours to retain their money, while those in one of the earn conditions begin with a small amount of money and have the opportunity to earn more every time they log on. When signing up for a game, participants are offered a security choice: whether to enable \emph{two-factor authentication} (2FA). Prior to making this choice they are shown explicit risks (which vary by game): risk of being hacked and amount of protection offered by adopting 2FA. If they are hacked -- probabilistically determined by a script regularly run on the system -- participants lose all of the money in their account. At the end of a game, participants receive the real monetary value of the amount left in their account. 

Participants in our experiment play the game twice, each time with a different condition (endowment/earn, hack risk, and protection offered by 2FA). We invited 150 AMT workers to play the first game. 
Based on the observations we record during game play, we model the factors that influence whether a participant enables 2FA and whether the decision was utility optimal (e.g., rational). We further examine the extent to which people make utility-maximizing decisions and how the optimization of choices varies based on effective risk and the endowment vs. earn condition.

We find that people's decisions to enable 2FA can be explained well (pseudo-${R}^2$ = 0.612) by their prior behavior (approximately 35\% of the variance), knowledge of costs (e.g., awareness of how long 2FA will take) (approx. 15\% of behavior variance), and to a lesser but still significant extent, explicit risk judgements and endowment effects (9\% to 15\% of variance, depending on their knowledge of other factors). Further, we find that both risk and endowment effects also relate to whether the decision to enable a behavior was utility-optimal. We find that those playing in the lowest-risk conditions are far less likely to make utility-optimal decisions than those in the highest-risk conditions. Those in the medium-risk conditions make a utility-optimal decision 58\% of the time the first time they play the game and 69\% of the time the second. 
Our results suggest that, in our experiment: 
\begin{itemize}
\item Users made rational security choices approximately 50\% of the time.
\item User tend to act in accordance with an anchoring effect~\cite{strack1997explaining}: they tend to stick with the first security decision they make. More general security tendencies (chosen password strength, security behavioral intentions) did not influence behavior.
\item Users are boundedly rational: they appear to incorporate knowledge about costs and explicit risks.
\item In higher-risk conditions, users enable security options more often and make rational decisions more often.
\item When protecting assets they already have, especially in highest-risk games, users tend to behave more rationally and more securely (endowment effect~\cite{knetsch1989endowment}).
\end{itemize}

Using this data, we can compute estimates of utility-optimizing behaviors for a given user based on users' typical costs (e.g., login speed, account balance, and earning potential) and estimated explicit risks (risk of hacking, protection offered by a behavior). As a tangible example, we present a thought experiment using our data to theorize about market gains and losses from different approaches to user security: encouraging all users to be secure, encouraging utility-optimal behavior, and not requiring end users to engage in security behaviors. We show that a ``one-size-fits-all'' emphasis on security can lead to significant market {\it losses} from needless user costs for negligible security utility, and identify hypothetical risk thresholds at which targeted encouragement of certain security behaviors for at-risk users can lead to market gains.

Consequently, we suggest a push toward personalized security recommendations (e.g., nudges) to guide users toward utility-optimal behaviors and allow them more autonomy to make decisions based on transparently communicated risks paired with a push for more data-driven research quantifying those risks. We conclude with further discussion of the implications of our findings for predicting user behaviors and improving market, or population, security through nudging and risk communication, and suggest directions for future work.
\section{Related Work}
\label{sec:related}
Below, we review prior work on the decision-making in security and privacy, focusing specifically on economics-based studies when available.

\textbf{Security Decision-Making.} Herley takes the first economic approach to examining end-user security decisions~\cite{Herley:2009:SLN:1719030.1719050}. In his theoretical paper, he provides a cost-benefit analysis of having end-users follow security advice. He argues that threats are so rare -- and suggested behaviors so ineffective -- as to make it logical for end users to never adopt security behaviors. Relatedly, B\"ohme and Grossklags approach user attention in security as a public good, that is, as an extremely scarce resource and examine how to facilitate optimizing rather than satisficing of user attention toward security warnings and privacy notices~\cite{bohme2011security}. They theoretically explore the effect of different user interface designs on the use of attention and recommend that user interactions should be rationed, rather than just user interfaces optimized, to maximize the use of users' privacy and security compliance budgets. 

The majority of security-related economics analysis of human behavior has focused on experts' (e.g., system or network administrators) responses to threat~\cite{johnson2010security,grossklags2008security,christin2004near,hota2015interdependent}%
, organizational responses to threat~\cite{miura2008security,gal2005economic,campbell2003economic}, or attackers~\cite{alpcan2006intrusion,liu2006bayesian,chen2009game,sallhammar2005using,manshaei2013game,shim2012crime}. For example, Grossklags and colleagues conduct a number of different studies applying game-theoretic models to examine expert behavior in attack/defense security games~\cite{grossklags2008secure,johnson2010security,grossklags2008security,christin2004near} finding evidence of unclear risks and bounded rationality for security professionals; and relatedly, Rounds et al.\ examined the impact of changes in security and reward on attacker responses, finding that as the value of an account increases, so too does the number of attacks on that account; and as the amount of security on an account increases so too does the number of attacks decrease~\cite{rounds2013experimental}.  These analyses of experts, however, provide limited insight into end-user behavior.

Christin et al.\ present one of the few empirical, economic-related, studies of security behavior: they incentivized users to execute arbitrary code~\cite{christin2011s}. They adjusted the incentives provided, finding that 43\% of participants were willing to run untrusted executables for \$1.00. Most prior work focused on end-user decision making has been more qualitative, examining people's security decisions through surveys and interview studies. These studies include explorations of why people say they update or do not update their software~\cite{mathur2016they,vaniea2016tales,wash2015too}%
, why users fall for phishing attacks~\cite{sheng2010falls,hong2012state}%
, why people bypass browser warning messages~\cite{harbach2013sorry,sunshine2009crying,felt2014experimenting}%
, and even preliminary support for boundedly-rational decision-making in cybersecurity based on post-event surveys with university students~\cite{aytes2004computer}. Findings have shown evidence of the importance of fatigue from too many security demands, unclear risks, and lack of clarity on the utility of behaviors for addressing risks.

The reasonable conclusion of many of these end-user studies is a push for better education of users to get them to take security risks seriously and enable security behaviors. Indeed, especially in the corporate realm, there may often exist an implicit assumption that more security is better and our goal should be to get all users to enable all security behaviors at all times~\cite{kirsch2007last,herley2017justifying}. Herley's work challenges this approach, arguing that security is \textit{never} better for the end-user, and at best benefits the corporations of whom they are clients while costing the individual~\cite{Herley:2009:SLN:1719030.1719050}. Underlying both approaches is also an implicit assumption that users have rational reasons for adopting or ignoring security advice, and are able to interpret and make reasonable decisions about risks if they are aware of them. In our work, we explore to what extent users are rational actors and we explore what factors, including implicit and explicit risks and costs, relate to the security decisions they make in our simulation. We also explore the impact of the ``users should always be secure'' and ``users should never be secure'' perspectives on the costs and utility of enabling 2FA for an entire population to better understand when, and whether, we should encourage end users to adopt security behaviors.

\textbf{Economics of Privacy.} Prior work on privacy has explored privacy as a good and theoretical privacy markets~\cite{hui2006economics,acquisti2016economics,farrell2012can,berthold2010valuating} as well as empirical examinations of users valuations of data~\cite{carrascal2013your,bauer2012value,friedman2007empirical,cvrcek2006study}, and especially their willingness to pay to protect private, digital information vs. willingness to disclose information for a payout~\cite{acquisti2013privacy,schudy2017you,schreiner2013willingness}. Acquisti and colleagues have found evidence of bounded rationality -- that is, semi-rational decisions in which people are only able to take into account a portion of the data (i.e., risks or evidence) in their decisions~\cite{Acquisti04:Privacy,acquisti2005privacy} and have found that difficultly identifying risks and benefits can lead to significant uncertainty that muddies privacy decision-making~\cite{acquisti2015privacy}. Closely related to our work, Tsai et al.\ explore what happens when making privacy tradeoffs more transparent, finding that when privacy information is surfaced to users in an online shopping scenario they are more likely to make purchases from a privacy-preserving website, even at a higher cost~\cite{tsai2011effect}. Beyond privacy, behavioral economics experiments similar to ours have been used effectively to model bounded-ly rational decisions in business and management settings, health, and purchasing~\cite{simon1979rational,diamond2012behavioral,hanoch2007bounded}.

Our work draws on findings from these studies to explore bounded-rationality in security and examine how security-tendencies such as password strength choices influence decision-making. We combine the individual-focused analyses presented by this prior work in privacy with the market-level analyses emphasized by security economics work, presenting a perspective on how and when we should emphasize security protection for end-users.

\section{Methodology}
\label{sec:methods}
Below we describe our experimental system, the experiments (e.g., games), participant recruitment, our analyses, and limitations. All procedures were approved by our institution's review board.

\subsection{System Design}
\begin{figure}
\small
\centering
\includegraphics[height=0.4\textheight]{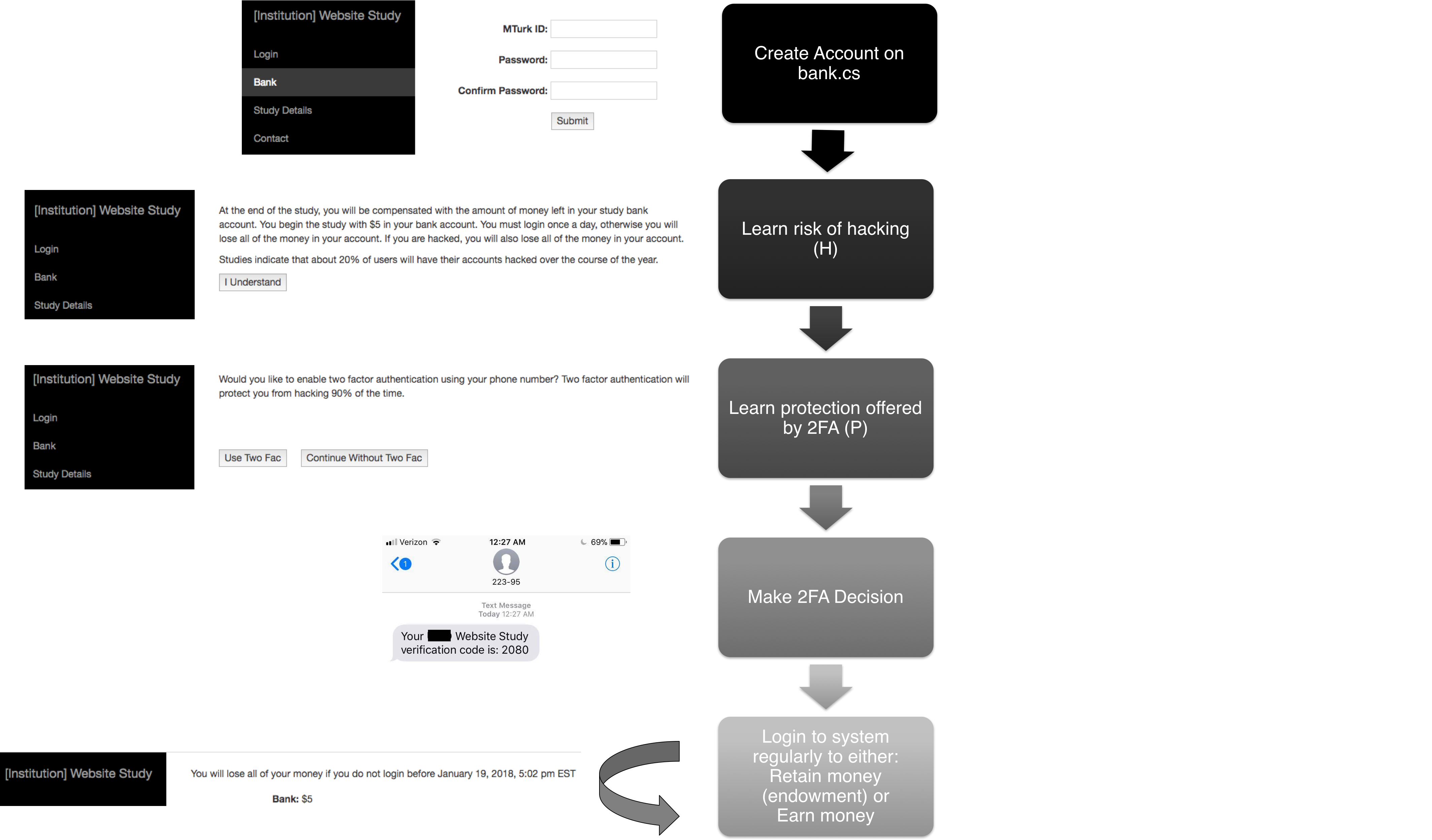}
\caption{Overview of the experimental system flow.}
\vspace{-0.5ex}
\label{fig:system}
\vspace{-2ex}
\end{figure}

We created the ``bank'' system to provide a similar UI flow to logging into a real bank account (Figure~\ref{fig:system} provides a system overview). The system URL was \url{bank.[institution domain]}, had an SSL certificate, and was thus displayed over HTTPS. 

When participants began the study, they were taken to a page that had two options: log in or sign up. After choosing the signup button they were taken to a page where they could create an account. The account creation page asked the participant for their Amazon Mechanical Turk (MTurk) ID and to create a password. There were no constraints for password creation, and the provided MTurk IDs were validated against the list of workers who accepted the task on MTurk. If the participant clicked the Login button, they would be taken to a page asking for their MTurk ID and their previously created password.

The next screen explained the study. In the study, we varied participants' game condition, which was a combination of experimental setting (Earn or Endowment), risk of hacking ($H$), and protection offered by 2FA ($P$). Those in the Earn Setting ($S:Earn$) started with \$1 in their account and could earn an extra \$1 per day up to a maximum of \$5. Those in Endowment Setting ($S:Endowment$) started with \$5 and would lose that money if they did not log in once per day. In addition to varying the conditions, we also transparently told participants their risk of getting hacked $H$.\ Hacking was experimentally constructed: we ran a script every evening, randomly selected a number, and ``hacked'' the participant (e.g., removed the money from their account) if the randomly selected number was less than their effective risk; hacking probabilities were not affected by how frequently participants chose to login. The participant's effective risk was defined as either $H$ (no 2FA selected) or $H*(1-P)$ (2FA selected), that is, the probability of being hacked modulated by the accepted protection. In the initial consent form and study instructions it was made clear to participants that the hacking was constructed and that there were not real hackers trying to break into their study accounts. 

Participants were randomly assigned one of three $H$ values: 20\% (a chance of hacking popularly cited, which we use as a benchmark~\cite{hackpercent1,nortonhack,hackpercent2}), 1\%, and 50\%. To convey this information, participants in $S:Earn$ saw the message shown in the second screen from the top in Figure~\ref{fig:system}. 
%
%
Those in $S:Endowment$ saw a very similar message, which replaced the second and third sentences in the message with: ``You begin the study with \$1 in your bank account. Each time you login (at most once per day) you will earn an additional \$1.'' 

Next, participants were offered the option to enable 2FA.  We chose 2FA as the exemplar behavior in our system because prior work finds that it is an ``in-the-middle'' behavior -- understood better and adopted more often than updating, but less so than strong passwords~\cite{redmiles2016learned}. Additionally, it was easier to create a simple, realistic system around a binary 2FA choice (enable/don't enable) than around other security behaviors such as updating or browser warnings.

We were explicit about the amount of protection offered by 2FA, just as we were explicit about the risk of being hacked. Participants were randomly assigned an arbitrarily chosen protection value, $P$, of 50\% or 90\%.  As defined above, that value was then used in the effective risk calculation that led to a chance of ``hacking'' in our experiment. Thus, their effective risk was $H(1-P)$. To communicate this, participants saw the following message: ``Would you like to enable two factor authentication using your phone number? Two factor authentication will protect you from hacking $P$ of the time.''

Those who selected 2FA went through SMS-based two-factor setup, where they received an initial text with a four-digit code that they needed to enter before proceeding.\footnote{While SMS-based 2FA has been shown to be insecure~\cite{reaves2016sending}, it is still the most broadly adopted 2FA standard. Additionally, SMS-based 2FA provides a simple, quantifiable privacy and time cost for our experimental scenario: app-based 2FA would complicate cost calculations, as we would not know if certain users already had the application we chose installed on their phones. For purposes of our experiment, the actual security offered by SMS-based 2FA in practice does not matter.} Then, on all subsequent logins, they received an SMS with a code and had to enter this code along with their password in order to log in. 

Finally, participants were shown their bank balance, and also reminded of when they must next log in, for $S:Endowment$, or when they could next earn more money, for $S:Earn$.

Upon subsequent logins, participants would see the previously described login page where they logged in using their MTurk ID, password, and, if they selected the option, 2FA. After logging in, the participant was directed to their bank balance page. At the end of the study, on their last login, they were shown a link to a Qualtrics survey which the user needed to complete in order to receive payment. The survey used two validated measures, SeBIS~\cite{Egelman:2015:SSW:2702123.2702249} and the Web-Use Skills Index~\cite{Hargittai:2012:SSM:2142476.2142477}, to measure the participant's general security behavior intentions and internet skill; we also collected their gender, age, and education level. 

\paragraph{Technical Details: Data Storage \& Measurement}
Our system was built using PHP 5.6.32. The 2FA validation was handled by the Authy
 API. All data was stored in a secure database. Only the hashed and salted password, not the cleartext password, was saved using the PHP password-hash function
 with PASSWORD\_DEFAULT. Passwords were verified using the PHP password-verify function
 .

To measure password strength, we used the data-driven neural-network password meter~\cite{ur2017design}. To record the signup and login times, we recorded the total number of seconds for which the 2FA screen was in focus. We began recording when the page was loaded and stopped recording focus time when the pages were submitted. That is, if a user went to the login page, stayed on that page for five seconds, checked their email for two minutes, and finally came back to the website and took another ten seconds to submit their login, we would have recorded a login screen time of fifteen seconds. Table~\ref{tab:variables} in Appendix~\ref{sec:appx:variables} provides a summary of all observed and self-report collected variables.

\subsection{Experimental Design}
%
We recruited 150 workers from Amazon Mechanical Turk (MTurk) to participate in two rounds of an experiment run on our bank system. We recruited only workers from the United States and those who had a 95\% approval rating or above, as prior work shows that these workers produce the highest quality results~\cite{peer2014reputation}. Our advertisement stated that participants would be evaluating a website for our institution and had the opportunity to earn up to \$5. In the first experiment, participants completed a consent form and were then redirected to create an account on the bank site. They progressed through the steps described above, and were assigned to one of eight experimental settings: H=1\%, P=50\%, Endowment and Earn (Low Risk); H=1\%, P=90\%, Endowment; H=20\%, P=50\% Endowment and Earn; H=50\% and P=50\% Endowment; H=50\% and P=90\% Endowment and Earn (High Risk). We selected a non-full-factorial design due to cost constraints; we selected the following conditions that would maximize our ability to draw conclusions about the effect of various risk levels as well as the endowment condition. The experiment ran for five days. All participants who started the experiment were paid \$0.50, and those who submitted their final surveys were paid the amount remaining in their bank account (ranging from \$0 through \$5.00).

Five days after the conclusion of Round 1 (RD1), participants received an email through MTurk inviting them to participate in the experiment again. They were told that things would work the same way as last time, but that they needed to create a new account. Participants were again randomly assigned to one of the eight experimental settings, with no relation to their RD1 settings. When participants finished Round 2 (RD2) they were not asked to take the survey again, but were again compensated as in RD1. Participants spent an average of a total of 142 seconds (Std. Dev =  34.57 (s)) logging in to RD1 and an average of a total of 158 seconds (Std. Dev =  30.40 (s)) logging into RD2, plus an average of 6 minutes completing the survey after the first game.

\subsection{Analysis}
\label{sec:meth:analysis}
Our analysis consists of three components: (1) modeling security behavior as a function of the variables measured and assigned in the games; (2) determining the degree to which participants' behavior was utility-optimal, that is ``rational,'' and modeling the factors correlating with rational decision-making; and (3) presenting a thought experiment showing how we could use this information to predict utility-optimal behaviors for MTurker's and how such predictions could be used to nudge security behavior in the future. We describe our specific methodology for each analysis in more detail in Sections~\ref{sec:results:behavior} to \ref{sec:thoughtexperiment}. Table~\ref{tab:regression:models} in the Appendix summarizes the models we construct.

\subsection{Limitations}
Behavioral economics experiments are subject to a number of limitations: participants may behave differently than in real life, our variables for the hack and protect percentages may be unrealistic, and 2FA may not be a representative security behavior. We have done our best to mitigate these limitations by setting our incentive in line with typical behavior economics incentives~\cite{kamenica2012behavioral} and by choosing hack and protection percentages both higher and lower than real values suggested by popular studies for the hack percentage, and we tried to pick protection values that would provide information about whether people feel differently about a nearly ``perfectly'' protective security behavior and a behavior that will only work half of the time. Finally, we chose 2FA as our exemplar behavior because prior work suggests it is a reasonable ``middle-of-the-road'' in terms of user understanding and adoption~\cite{redmiles2016learned}. 

It is also possible that the source of our sample, Amazon Mechanical Turk, and the fact that our sample is not representative of the demographics of the US may have biased our results (our participants are younger and more educated than the U.S. population, as is typicaly of AMT samples~\cite{ross2010crowdworkers}); additionally it is possible that those participants who lost their money due to hacking or due to forgetting to log in would be far less likely to return their surveys or return for the second round. We chose to use Mechanical Turk because census-representative survey platforms do not allow for the deployment of experiments such as ours, and we experienced low drop out rates (17\% RD1, 15\% RD2) suggesting that lack of compensation for participants who were hacked or forgot to log in was not a significant source of bias. 
\section{Participant Descriptives}
125 of the original 150 participants completed RD1. 107 of these RD1 participants finished RD2. The demographics of the two groups were nearly identical. Our samples were skewed slightly male (54\% in R1, 55\% in R2). Our participants were more educated than the general population, with 51\% and 50\% in each round, respectively, holding a college degree or above as compared to 28\% of the U.S. population~\cite{census:acs}. Finally, our samples are both skewed heavily younger than the general population, with 70\% of our participants under 40 years old compared to 40\% of the general population. Table~\ref{tab:demo:census} in the Appendix shows the demographics of our samples. 

Our RD1 participants had a mean internet skill score of 4.06 on the Web-Use Skills Index~\cite{Hargittai:2012:SSM:2142476.2142477}; our RD2 participants had a mean internet skill score of 4.016 (this difference was not significant by a Mann-Whitney U test). Our RD1 participants had a mean SeBIS score~\cite{Egelman:2015:SSW:2702123.2702249} of 3.17; RD2 participants had a mean SeBIS score of 3.16 (the difference in SeBIS scores across rounds was not significant by a Mann-Whitney U test).

%
%
%
\section{What Factors Drive People's Security Decision-Making?}
\label{sec:results:behavior}
\textbf{Analysis Approach.} To model our participants' behavior, we constructed binomial logistic regression models where enabling 2FA was the outcome variable. For modeling 2FA enablement in RD1, we included the following input variables: the participant's gender, age, educational level, internet skill, security behavior intention and password strength, as well as the protect and hack percentages presented, the setting (earn or endow), and two-way interactions between the latter three factors. We used backward selection based on the Bayesian Information Criterion to select the model of best fit~\cite{spiegelhalter2002bayesian}. We report model fit using McFadden's Pseudo-${R}^2$~\cite{cameron1997r}, approximating the percent of variance explained. 

For modeling round two behavior, we construct three models: one with only RD1 behavior as the input variable; one with RD1 behavior, RD1 signup time (not including time spent on 2FA), and RD1 mean login time (not including time spent on 2FA); and one with RD1 behavior and times as well as RD2 equivalents of the factors found to be relevant in the RD1 2FA enablement model (a subset of: hack and protection percentages, demographics, password strength, the setting, and relevant interactions) as the input variables. We chose to construct these models in order to examine the relationship between security decision-making and anchoring effects,\footnote{The anchoring effect is a commonly shown psychological
effect in which humans tend to stick with the first decision
they make or choice they are offered~\cite{strack1997explaining}.} participants' costs (e.g., tendency to login or signup slowly), and the relevant factors identified in RD1 (our ``blind'' experiment, in which we do not yet know participants' typical tendencies or costs). We do not include whether the participant was hacked or lost money in RD1 because fewer than 10 such participants returned to play in RD2, a quantity insufficient to provide robust regression results. We compared the fits of our three models using likelihood ratio tests~\cite{bentler1980significance}. 

We report the results of all models, including log-adjusted regression coefficients (odds ratios), 95\% confidence intervals for those odds ratios, p-values (with a significance threshold of 0.05), pseudo-${R}^2$, and the results of model comparison tests, when appropriate.

\begin{figure}
\footnotesize
\centering
\includegraphics[width=\columnwidth]{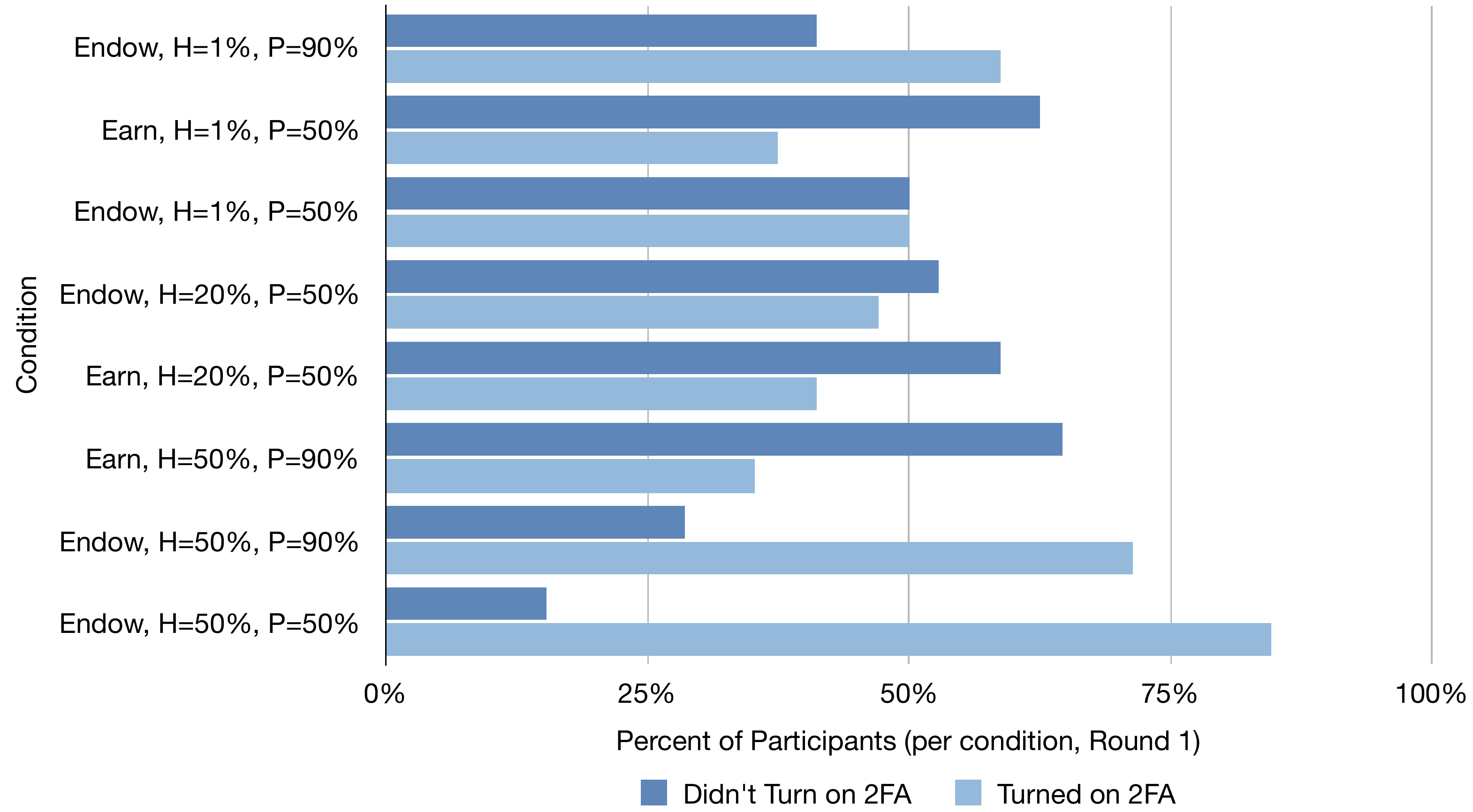}
\vspace{-3ex}
\caption{Percent of participants who enabled 2FA in each of our eight experimental conditions in RD1.}
\label{fig:choices:rd1}
\vspace{-1ex}
\end{figure}

\textbf{Factors that Drive 2FA Decisions.} 51\% of participants in RD1 and 56\% in RD2 chose to enable 2FA. These decisions varied by condition, as shown in Figure~\ref{fig:choices:rd1}. 

\begin{table}
\centering
\footnotesize
\begin{tabular}{lrrr}
 \hline
 Variable & O.R. & C.I. & p-value \\ 
 \hline
 $Endowment$ & 2.32 & [1.44, 3.76] & $<$0.001* \\ 
 $H$ & 2.31 & [1.22, 4.38] & 0.011* \\ 
 $P$ & 1.46 & [1.22, 1.97] & 0.043* \\ 
 $Endowment$:$P$ & 3.61 & [1.35, 9.67] & 0.012* \\ 
 \hline
\end{tabular}
\vspace{-0.5ex}
\caption{Logistic regression results for 2FA behavior in RD1. O.R. is the log-adjusted regression coefficient (odds ratio), CI is the 95\% confidence interval of that odds ratio, and p-values marked with * are significant at the 0.05 level. Endowment is a boolean variable for whether the participant was in one of the endowment conditions, $H$ and $P$ are numeric, and ':' indicates an interaction variable.}
\label{tab:rd1:regression}
\vspace{-2ex}
\end{table}

We model RD1 decisions to enable 2FA as a function of attributes of the respondent (gender, age, education, internet skill), and security practices or savviness (security behavior intention, password strength), and risks and conditions assigned to the participant ($H$, $P$, $S$), as well as interactions between these risks and settings: interaction between endowment (T/F) and $H$ and interaction between endowment and $P$. We find that the model of best fit retains only the risk, protection, and setting factors, as well as the interaction between setting and protect percentage (Table~\ref{tab:rd1:regression}). Those in the endowment conditions are 2.3 times more likely to enable 2FA, in line with endowment effects observed in other fields~\cite{knetsch1989endowment,acquisti2013privacy}. 

Those who are shown a higher $H$, that is told that research shows a higher chance of account hacking, are more likely to enable 2FA. Those shown a higher $P$, that is, those told 2FA will be 90\% rather than 50\% effective, are also more likely to enable 2FA, while those in a condition that involved endowment and a higher $P$ are even more more likely to enable than just accounting for $P$ or the endowment setting separately would suggest. This model explains 14.7\% of the variance in the RD1 data.\footnote{We note that this is likely a \emph{conservative} interpretation of the effect; indeed, psychologists have suggested that $R^2$ values may appear deceptively low when modeling singular events (e.g., decisions) that tend to accumulate over time, and that the true explanatory power of the significant variables in these cases may be stronger than a singular measurement indicates~\cite{abelson1985variance}.}

As described above, we model RD2 decisions with three different sets of models. The model with just RD1 behavior as an input factor indicates that those who enabled 2FA in RD1 are 83 times more likely to do so in RD2 (Appendix Table~\ref{tab:rd2:rd1:regression})
, in line with the expected anchoring effect. This model explains 35.3\% of the variance in our RD2 data. Next, we added RD1 costs (e.g., RD1 signup and login times) to the model to understand whether RD2 behavior was explained just by repeating RD1 behavior or also by learning from the costs of RD1 behavior. 
The model of best fit retained all three factors, and we see that both login time in RD1 and RD1 behavior are significant in the model, with those who enabled in RD1 being more likely to enable in RD2 and those who took longer than others to login in RD1 being 6\% less likely to enable 2FA in RD2 (Appendix Table~\ref{tab:rd2:threefac}). 
This model explains the data significantly better than the prior model ($p=0.012$) with pseudo-${R}^2$=0.522. Finally, we built a third model that included these three RD1 factors as well as the experimental settings retained in RD1. This final model shows only that enabling 2FA in RD1 is significantly related to enabling 2FA in RD2, however, we find that this final model with the experimental variables fits significantly better than the prior model ($p=0.040$) and explains a total of 61.2\% of the variance (Table~\ref{tab:rd2:final:regression}), suggesting that the costs and experimental settings are explanatory of behavior but not as strongly as are anchoring effects. 

\begin{table}
\centering
\footnotesize
\begin{tabular}{lrrr}
 \hline
 Variable & O.R. & C.I. & p-value \\ 
 \hline
 R1 2FA Decision & 32.66 & [2.19, 486.54] & 0.011*\\
 R1 Login Times & 0.81 & [0.6, 1.11] & 0.190 \\ 
  R1 Signup Time & 0.90 & [0.69, 1.16] & 0.411 \\ 
 $Endowment$ & 1.26 & [0, 547.12] & 0.094 \\ 
 $P$ & 17.09 & [0.04, 144.74] & 0.357 \\ 
 $H$ & 72.88 & [0.57, 373.07] & 0.063\\
 $Endowment$:$P$ & 0.99 & [0.91, 1.08] & 0.822 \\ 
R1 2FA:Login Times & 1.12 & [0.81, 1.56] & 0.496 \\ 
R1 2FA:Signup Time & 2.41 & [0.26, 22.02] & 0.435 \\ 
 \hline
\end{tabular}
\vspace{-0.5ex}
\caption{Logistic regression results for 2FA behavior in RD2 modeled as a function of RD1 2FA decision, RD1 costs (login and signup times), and RD2 experimental settings. McFadden Pseudo-$R^2$=0.612}
\label{tab:rd2:final:regression}
\vspace{-2ex}
\end{table}

\section{Do People Make Rational Security Decisions?}
\label{sec:results:rational}
\textbf{Analysis Approach.} To examine whether participants made utility-optimal decisions and the factors related to the rationality of those decisions, we considered participants to have made a \textit{utility-optimal} decision in the following way: it is utility-optimal to enable 2FA if the cost of doing so is less than the expected utility that would be gained from 2FA. For those who enabled 2FA in either round, we compute the cost of using 2FA for an individual user as the time it cost them to sign up (in hours) plus the sum of the time it cost them to log in each time (in hours) times their hourly wage:
 $C_{\mathit{2fa}}  = (T_{\mathit{signup}} + \sum T_{\mathit{login}})*wage_{\mathit{mturk}}$. 
To compute the value of our participants' $wage$ we used the data collected in the Pew 2016 survey about MTurk workers~\cite{pew:2016} to compute a population-based estimate of \$4.97/hr (calculations in Appendix~\ref{sec:appx:wage})
. For those who did not use 2FA, and in any analysis in which we use the full dataset (that is the data of those who enabled and did not enable 2FA), we estimate the cost of 2FA. A sensitivity analysis reveals a linear pattern in estimations: considering 2FA cost as the mean results in an estimate of more rational choices than considering 2FA cost as $2\times$ or $3\times$ the mean. To generate conservative estimates of rational behavior we continue our computations by estimating 2FA cost for those who did not enable as $2\times$ the mean.

We compute the utility of using 2FA as the potential loss (maximum amount they could earn times hack percentage) times the protection gained by using 2FA ($P$):
 $U_{\mathit{2fa}}  = P[(H)*{\mathit{Max}}_{\mathit{bank}}]$. 
Where ${\mathit{Max}}_{\mathit{bank}}$ was \$5 for our experiment, and is the value of the money in an MTurker's account for the third analysis.

Below we examine whether 2FA users made utility-optimal decisions. We also examine the proportion of utility-optimal decisions made by our sample as a whole using the mean cost of enabling 2FA (computed from those who did choose to use it). To compare the strength of the associations between the decisions made in each round, we use a ${\chi}^2$ test with Cram\'{e}r's V
 as our measure of effect size and use Holm-Bonferroni correction to account for multiple comparisons
 .
 
Finally, we model whether participants' decisions were utility-rational using regression models. We construct these models using a similar approach to that described in Section~\ref{sec:results:behavior}, with RD1 utility-optimal decision-making being modeled as a function of the participant's gender, age, educational level, internet skill, security behavior intention and password strength, as well as the protect and hack percentages presented, the setting, and two-way interactions among the latter three factors. RD2 utility-optimal decision-making (True or False) was modeled with two models: the first was a single-factor model containing only whether the RD1 decision was utility-optimal, and the second included the same factors as in RD1 as well as the RD1 decision. We performed backward selection on this second model to achieve a model of best fit for RD2.

\textbf{Were Participants' Decisions Rational?} We find that, following the approach described above, 48\% of all participants made utility-optimal decisions in RD1 and 58\% did so in RD2.

\begin{figure}[t]
\footnotesize
\centering
\includegraphics[height=0.25\textheight]{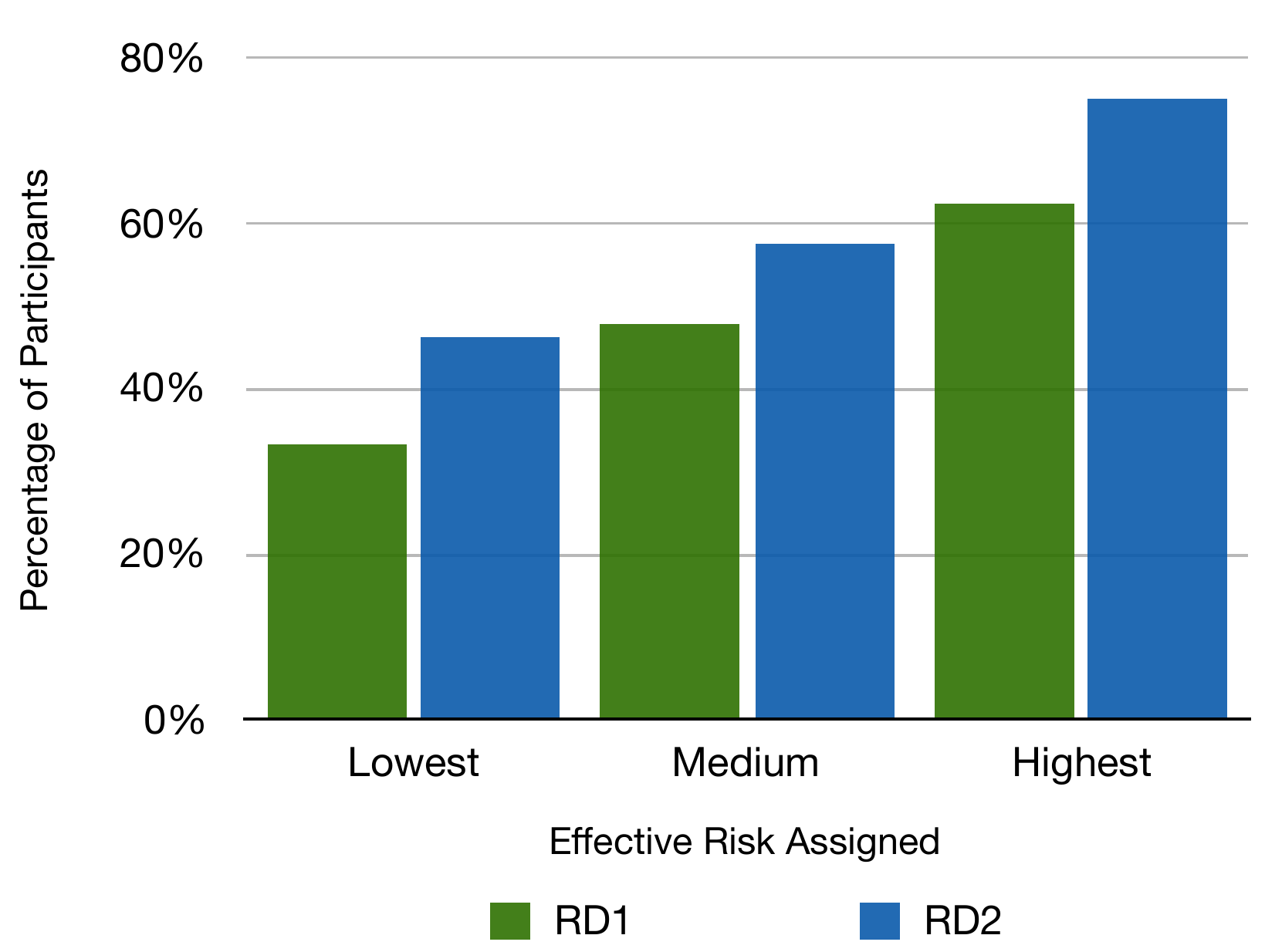}
\vspace{-0.5ex}
\caption{Percentage of participants in each round and with a given risk setting who made a utility-optimal decision about 2FA.}
\label{fig:utility:comp}
\vspace{-2ex}
\end{figure}

We also consider behavior at the extremes: that is, the lowest and highest risk conditions, as well as in the middle. 
%
We find that, in RD1, 33\% of participants in the lowest-risk settings (e.g., did not enable 2FA), 48\% in the medium risk, and 63\% in the highest risk settings make a utility-optimal decision (e.g., enabled 2FA~\footnote{In our particular experiment, using 2FA was always utility-optimal in the highest risk setting and never in the lowest.}). We observe a learning effect (${\chi}^2 = 21.226$, df$=2$, $p<0.001$, V = $0.578$ (medium)) with 58\% of all participants in the medium-risk experiments making a rational decision in RD2, 46\% in the lowest-risk settings and 75\% in the highest-risk settings. Figure~\ref{fig:utility:comp} compares the change in utility-optimal behavior by risk level for the rounds.
%

\textbf{Modeling utility-optimal decision making.} We wanted to understand how these different factors, as well as participant factors, interacted to potentially explain whether participants' make utility-optimal decisions.

We first model whether participants made a utility-optimal 2FA decision in RD1 in the same way as we modeled general decision-making for RD1 in Section~\ref{sec:results:behavior} above. The model of best fit retains only the hack percentage, setting, and internet skill factors (Appendix Table~\ref{tab:rd1:utility:regression}; 
pseudo-$R^2$=0.141). Those in the endowment setting are 25\% more likely to make a utility-optimal choice, while those with higher internet skill are 15\% more likely to do so. Further, we see that those who saw a higher hack percentage are more likely to make a \textit{utility-optimal decision} in RD1.

For RD2, we find that a single-factor consisting of only whether the RD1 decision was utility optimal is not significant. After backward selection, the multi-factor model of best fit retains SeBIS, the risk and protection factors, endowment, and interactions between risk and endowment and protection and endowment (Appendix Table~\ref{tab:rd2:utility:regression}; psuedo-$R^2$=0.078).  We see that those with higher SeBIS scores are nearly four times more likely to have made a utility-optimal decision in RD2, and (similar to RD1) those who saw a higher hack percentage are 21\% more likely to make a utility-optimal decision.
\section{Thought Experiment: When Does Security Benefit the Market?}
\label{sec:thoughtexperiment}
In order to examine the impact of different models of user behavior on market costs and utility from security, we present a simple thought experiment regarding the implementation of optional 2FA on the MTurk platform.

\textbf{Approach.} Suppose that MTurk decided to offer a 2FA option to workers. Workers who enable 2FA then must use 2FA every time they log in, exactly as in our experiment. We know from prior work~\cite{pew:2016} that 63\% of MTurkers log in daily, 32\% multiple times a week (for the purposes of our thought experiment, we say $3\times$ per week), 3\% log in weekly, and the rest log in less than once a week. Using costs (as defined in Section~\ref{sec:results:rational}) from our experiment, this means 2FA would cost: $C_{2fa}  = 0.065*{wage}_{mturk} + 0.039 * N(logins)*{wage}_{mturk}$. 
Using this equation, we see that each week using 2FA would cost the daily login group \$0.34, the $3\times$ a week login group \$0.18, and the weekly login group \$0.10.

Suppose also that MTurkers transfer the money they earn out of their accounts once per week, so the value of their bank (e.g., the money they stand to lose if their account is hacked) is the value of their weekly earnings. Not all MTurkers earn the same amount on MTurk; prior work shows that 17\% earn all of their income on MTurk, 8\% earn most (75\%), 6\% earn half (50\%), 15\% earn some (25\%), and 53\% earn very little (5\%) of their income on the platform. We estimate, based on reports from the population that those who earn all or most of their income on MTurk make \$38,555 per year or \$741 per week in total, with either 100\% or 75\% of that income coming from MTurk. Those who earn half or less of their income on MTurk earn a higher overall income of \$54,300 or \$1044 per week, but only 5\%-50\% of that income comes from MTurk. The header of Table~\ref{tab:thoughtexp} shows the estimated weekly earnings -- that is, the amount of money an MTurker would stand to lose -- for the different categories of earnings.

Finally, we estimate the utility gained from 2FA for some people at an arbitrarily selected 1\% risk of hacking over the course of the year and for others at an arbitrary 20\% risk of hacking over the course of the year. 
We suppose in this thought experiment that 25\% of people might choose weak passwords, thus having a hypothetically supposed 20\% risk of hacking over the course of the year. The remaining 75\% of the population we theorize have a 1\% risk of hacking. Thus, the utility of 2FA in a given week would be calculated as follows (with a conservatively assumed 50\% protection value): $U_{2fa}  = P(H*Income)/52$. We again consider enabling 2FA to be utility optimal in cases where the utility is greater than the cost. 

It is important to note that we choose fairly arbitrary values for this thought experiment, so the results should be considered as an example rather than taken literally. However, we believe this example (as with prior theoretical work that uses a similar approach~\cite{Herley:2009:SLN:1719030.1719050}) provides a useful perspective for thinking carefully about when and how to advise users.  
%
%

\textbf{Results.} We find that under the above assumptions, for all but those who earn the least on MTurk (\$52 per week) those who are at a 20\% risk of hacking and who log in more than once a week should always enable 2FA. Those that log in once a week could enable 2FA or not enable 2FA with no discrepancy in outcome, as the cost of enabling and the utility gain are equal. For those who earn the least on MTurk, it is never utility optimal to enable 2FA. 

\begin{table}[t]
\centering
\small
\begin{tabular}{lrrr}
{\bf Approach} & {\bf Costs} & {\bf Utility} & {\bf Loss/Gain}\\
\hline
All Secure & \$275 & \$148 & (-) \$126 \\
Utility Optimal & \$32 & \$128 & (+) \$96\\
No Security & \$0 & \$0 & ${\$0}^*$\\
\hline
\end{tabular}
\vspace{-0.5ex}
\caption{Estimated costs, utility, and loss or gain from each of the security approaches. Figures are per 1000 MTurk users. *We use the \protect\cite{Herley:2009:SLN:1719030.1719050} model, in which users sustain no losses from hacking, to compute this value. If users sustained losses, the value would instead be \$266.}
\label{tab:population}
\vspace{-4ex}
\end{table}

Based on these calculations, we compute the cost and utility, and the overall gain or loss, from the three different recommendations for user security: users should always take all security precautions, users should never take security precautions, and users should take utility-optimal security precautions. Table~\ref{tab:population} summarizes the results per 1000 MTurkers. We see that utility-optimal behavior leads to a gain of \$96 per 1000 MTurkers, while an ``enable security always'' solution leads to a cost of \$126 per 1000 MTurkers, and a ``never enable security'' solution leads to a gain of \$0. Estimates suggest that there are 500,000 MTurkers active at any given time~\cite{pew:2016}. Using this estimate, requiring all MTurkers to use 2FA would cost the market \$63,606, while providing MTurkers with nudges for suggested behavior that is utility-optimal would lead to a gain of \$47,865 if all MTurkers behaved optimally and \$23,932 if 50\% chose a utility optimal behavior. Of course, the current option of not presenting 2FA leads to no gain or loss.

More generally, assuming that 2FA offers protection only 50\% of the time, we find that for those who have \$741 in their accounts (e.g., earn \$741 weekly and withdraw their income once per week) there need only be a 1.5\% chance of their account being hacked over the course of the year for there to be utility gain from 2FA. For those who login weekly, a 2.7\% risk for those who login three times a week, and a 4.9\% risk for those who login daily. Similarly, for those who earn half their income from MTurk (\$522) and login weekly there need only be a 2.2\% chance of their account getting hacked over the course of the year for there to be utility gain from 2FA. (These figures include only monetary losses, not time losses or emotional consequences from cleaning up from hacking; including these costs would cause 2FA to become utility optimal at even lower risks.) On the other hand, for those who earn very little (or keep very little money in their MTurk accounts), the risks must be very large --- 68\% for daily logins or 21\% for weekly logins --- for 2FA to be utility optimal. Table~\ref{tab:thoughtexp} in the Appendix summarizes different risk and income levels at which 2FA may be utility optimal.


%
\section{Discussion}
\label{sec:discussion}
\textbf{Make Risks, Costs, and Benefits Transparent.} Overall, we find that explicitly stated risks and utilities (e.g., benefits) 
appear to explain between 9\% (Round 2) and 15\% (Round 1) of the variance in our participants' behavior. While participants lean heavily on anchoring effects (e.g., their typical practices) and to a lesser extent their knowledge of their own costs (e.g., how long it takes them to perform a login) when settling on a behavior, the additional contribution of risk and utility variables to the variance explained suggests that \textit{numeric} descriptions of risk and benefit are at least partially informing users' security decision making. This finding agrees with prior findings in privacy that showed people were able to pick between different platforms verbally described to be more or less privacy protective~\cite{tsai2011effect}. These findings consequently offer support for making risk and benefit information more transparent to users. Rather than telling people they ``should adopt behavior X to be more secure'' we may wish to give them more autonomy over their choices. 

Inherently, making risks (or benefits from a given behavior) more transparent requires us to know what the risk or benefit of a particular attack or behavior is to the user. Measuring this risk effectively is a challenging open problem~\cite{herley2017justifying}. Recent work, such as that conducted by Bilge et al.\ and Thomas et al.~\cite{46437,BilgeHD17}, pushes for more scientific and measurement oriented approaches to assessing the potential for an attack or the benefit of a security behavior. Our results provide additional motivation for this research direction, and present a parallel direction for future work exploring how best to convey risk to users (e.g., exploring how they interpret and respond to different textual and numeric estimates of risk), as well as investigating how explicit risk communication may factor into existing legal and policy frameworks around digital security and privacy disclosure.

\textbf{Recommend Utility-Optimal Behaviors.} In our experiment, being able to compute a cost for an individual's security behavior and their expected utility from that behavior allows us to predict what users should do in a particular security decision-making scenario. A similar principle could be used in a real-world system to develop personalized security recommendations to guide users toward utility-optimal behaviors, rather than always trying to encourage adoption of security behaviors. That is, by measuring users' typical time to complete certain behaviors (e.g., login), their implicit risk (e.g., password strength, security settings), and the history of attacks on a given platform for users with a similar risk profile, systems may be able to offer users personalized, suggested behaviors. 

For example, a platform could display a message similar to the following on login or when the user is confronted with a security decision: ``Our personalized security analysis system suggests that two-factor authentication may be a good fit for securing your account. We expect to take only an additional 30 seconds of your time.'' While prior work has found that encouraging users to adopt behaviors that their friends have adopted~\cite{Das:2014:ISS:2660267.2660271} can succeed, doing so may not always be of benefit to the user. A more carefully personalized system that takes into account the utility-cost ratio for the user, as well as other factors unexplored here, such as risk aversion, may offer a more trustworthy set of suggestions that are respectful of user time and reduce their security burden.

\textbf{Security-Savvy Not Necessarily Safer.} Our findings reveal that security-savvy, measured by password strength and SeBIS score, had no impact on choice to enable 2FA, and an inconsistent impact on rationality of decision. We had anticipated a relationship between 2FA use and password strength (or security behavior intention), either that stronger passwords (or more security intention) would indicate higher paranoia that would correlate with more 2FA use, or that those who picked weaker passwords would be more likely to enable 2FA, suggesting a tradeoff between the two risk-reducing behaviors. Instead, we find only a relationship between explicitly stated risks (e.g., the chance of getting hacked that we conveyed to them). This suggests that users treat different security behaviors independently without considering them as a package that influences  overall risk or may be unaware of the relationship between their own choices (e.g., the strength of the password they pick) and their risks. If so, they may have difficulty making rational choices.

Thus, future work may wish to explore users' level of understanding of the risks of their security choices, how they think these risks interact, and investigate how to correct resultant misperceptions. Additionally, we may wish to understand the efficacy of translating user choices into explicit risk assessments; for example, instead of presenting a scale of password strength from weak to strong, present password strength in terms of risk of account compromise. 

\textbf{Future Work.} The work presented here takes a first step toward measuring the rationality of user decision-making, assessing the contexts in which more and less rational behavior occurs, and understanding how risks, costs, and user factors influence decisions in security. Our experimental scenario presents the simplest of situations: with explicit, monetary, easily characterizable costs and risks. Future work may wish to explore rationality of decision making around data or other non-monetary assets and the impact of non-numeric risk statements, or explore other cognitive biases, e.g., present bias~\cite{Kleinberg16:Planning,Gravin16:Procrastination,Kleinberg17:Planning}.  Learning and/or comparing against other models of bounded rationality is of practical interest~\cite{Wright14:Level0,Hartford16:Deep}.
Additionally, our experiment only explores users' choice to enable 2FA, and indirectly their choice of password strength. Future work should validate, or invalidate, these results with other security behaviors. 

Finally, our results show that self-report measures of security behavior intention and internet skill were not related to 2FA enablement behavior. However, we see evidence that skill and/or intent may be related to the rationality of decision-making. This suggests that self-report measures may be ineffective at predicting or assessing behaviors --- a finding supported by extensive work in other fields~\cite{tourangeau2007sensitive,fendrich1994diminished,redmiles2017summary} --- but may be useful in assessing more deeply underlying constructs that are related to rationality. Future work should seek to further explore the utility of self-report measures for assessing security behavior, and whether experimental systems such as that presented here more, less, or equally as useful as self-reporting for measuring real-world behavior. 
%

%
\section{Conclusion}
\label{sec:conclusion}
In this work we explore the rationality of decisions made by end-users when faced with security choices. We validate that, at least in our relatively realistic experimental scenario, end-users behave in accordance with bounded rationality. That is, they are able to take into account some, but not all, risk and context factors and make rational (e.g., utility optimal) decisions in more than 50\% of cases. We find, further, that we are able to explain user decision making well (McFadden Pseudo-$R^2$=0.61) by accounting for participants' knowledge of costs and risks as well as accounting for the known phenomena of endowment and anchoring effects. 

We find that our participants made utility-optimal decisions more often when faced with higher risks. While perhaps encouraging for corporate high-risk scenarios, this finding also suggests a challenge for day-to-day security, as many of the risks end users confront in daily digital life are less transparent, less monetarily linked, and relatively small. Thus, future work may wish to explore how rationality is affected by different methods of communicating risk, less tangible consequences than the monetary incentives provided in our experiments, and even smaller risks. 

Using our data to compute utilities and costs for 2FA on the Amazon Mechanical Turk (MTurk) platform, we show that a ``one-size-fits-all'' approach encouraging all users to engage in all security behaviors at all times can lead to significant market and individual losses. However, we also show that there are relatively low risk thresholds at which 2FA becomes utility optimal, leading to significant market gains. This underscores the importance of quantifying the risk and utility of security behaviors through data-driven research~\cite{46437,BilgeHD17} in order to ensure that risks, costs, and benefits are accurately understood. 

Overall, our work supports a nuanced model of the ``human-in-the-loop'' who is able to somewhat take into account explicit risks and personal costs, in addition to relying on typical tendencies (e.g., anchoring effects), to make frequently rational decisions but who struggles to identify less obvious risks such as those incurred from weak passwords. This argues for personalized security-behavior recommendations for users tailored based on their costs (e.g., login times), risks (e.g., password strength), and value of their account (e.g., measured through the amount of money stored), especially in cases where the optimal behavior for them goes against their typical tendencies. If accurate, such personalized recommendations could provide security benefits while helping to avoid large market and personal costs from wasted time and effort on non-utility-optimizing security behaviors.


\section*{Acknowledgements}
Elissa Redmiles acknowledges support from the National Science Foundation Graduate
Research Fellowship Program under Grant No. DGE 1322106 and a Facebook Fellowship.
%
\bibliographystyle{ACM-Reference-Format}
\bibliography{bank,refs,new_refs}
\clearpage
\section*{Appendix}
\section{Participant Demographics}
\label{sec:appx:demographics}
Table~\ref{tab:demo:census} describes the demographics of the MTurk users who participated in both rounds of our experiment. The table compares their demgoraphics to those of the U.S. Census~\cite{census:acs}. Note that our participants are more educated (51\% and 50\% of the participants in the rounds, respectively, hold a B.S. compared to 28\% of the U.S. population) and younger (74\% and  73\%, respectively, are under the age of 39 compared to 40\% of the U.S. population). 

\begin{table}[h]
\centering
\small
\begin{tabular}{lrrr}
\toprule
{\bf Metric} & {\bf Round 1} & {\bf Round 2} & {\bf Census}\\
\midrule
Male & 54\% & 55\% & 49\%\\
Female & 45\% & 44\%& 51\%\\
\\
HS or Less & 13\% & 13\%& 28\%\\
Some College & 36\% & 37\% & 31\%\\
B.S. or Above & 51\% & 50\% & 28\%\\
\\
18-29 years & 31\% & 32\% & 23\%\\
30-39 years & 43\% & 41\% & 17\%\\
40-49 years  & 15\% & 15\% & 17\%\\
50-59 years & 8\% & 9\% & 18\%\\
60+ years &2\% & 2\% & 25\%\\
\bottomrule
\end{tabular}
\caption[]{Demographics of participants in our sample. Some percentages may not add to 100\% due to item non-response. Census statistics from the American Community Survey~\cite{census:acs}.} 
\label{tab:demo:census}
\end{table}
\clearpage
\section{Variables Measured in the Experiment}
\label{sec:appx:variables}
In Table~\ref{tab:variables} we summarize the variables measured in our experiment and subsequently included in our regression analyses. We outline the variable, method of measurement, and data source (either observations from game play in our system or the self-report survey administered to participants after RD1). 

\begin{table}[h]
\centering
\footnotesize
\begin{tabular}{lll}
 \textbf{Variable} & \textbf{Method of Measurement} & \textbf{Data Source} \\ 
 \hline 
2FA Decision & Enabled/Didn't Enable & System\\
&&\\
Password Strength & Data-Driven Neural-Network Password Strength~\cite{ur2017design}. & System\\
&&\\
Signup \& Login Times & Seconds each screen was in focus on participant's browser. & System\\
&&\\
Internet Skill & Validated measure of web-use skills~\cite{Hargittai:2012:SSM:2142476.2142477}. & Survey\\
&&\\
Security Behavior Intention & Validated measure of security behavior intention~\cite{Egelman:2015:SSW:2702123.2702249}. & Survey\\
&&\\
Gender & Male/Female/Other & Survey\\
&&\\
Education & HS or less/Some college/Bachelors or above & Survey\\
&&\\
Age & Numeric & Survey\\ 
  \hline
\end{tabular}
\caption{Variables measured in our experiment.}
\label{tab:variables}
\end{table}
\clearpage
\section{Computation of MTurk Wage}
\label{sec:appx:wage}
Here we outline our computation of the average MTurk wage for the population. The figures in the computation were drawn from a Pew Research survey of Amazon Mechanical Turk workers, which provides wage estimates for different portions of the MTurk population~\cite{pew:2016}.

\begin{align*}
wage_{mturk}  &= 0.01(\$12) + 0.03(\$10)+0.04(\$8)\\
&\qquad+0.07(\$7)+0.11(\$6)+ 0.2(\$5)+0.52(\$4) \\
&=\$4.97/hr\\ \vspace{-1ex}
\end{align*} 
\clearpage
\section{Regression Models and Tables}
\label{sec:appx:regression}
\subsection{Models Constructed}
Table~\ref{tab:regression:models} summarizes the regression models that we constructed during our analysis and references the tables in which the results of these models are contained.

\begin{table}[h!]
\begin{tabularx}{\textwidth}{lXr}
Outcome Var & Input Vars & Result Table\\
\hline
R1 2FA Decision & S, H, P, H:S, P:S, age, gender, education, IS, pwd. strength, SeBIS & Section~\ref{sec:results:behavior} Table~\ref{tab:rd1:regression}\\
R2 2FA Decision & R1 2FA Decision & Appendix Table~\ref{tab:rd2:rd1:regression}\\
R2 2FA Decision & R1 2FA Decision, R1 Login Time, R1 Signup Time, interactions between times and 2FA decision & Appendix Table~\ref{tab:rd2:threefac}\\
R2 2FA Decision & R1 2FA Decision, R1 Login Time, R1 Signup Time, interactions between times and 2FA decision, S, H, P, P:S & Section~\ref{sec:results:behavior} Table~\ref{tab:rd2:final:regression}\\
R1 2FA Rationality & S, H, P, H:S, P:S, age, gender, education, IS, pwd. strength, SeBIS &Appendix Table~\ref{tab:rd1:utility:regression}\\
R2 2FA Rationality & S, H, P, H:S, P:S, age, gender, education, IS, pwd. strength, SeBIS &Appendix Table~\ref{tab:rd2:utility:regression} \\
\hline
\end{tabularx}
\caption{Regression models constructed. S stands for setting (Earn or Endowment, with Earn as baseline); H is a numeric variable for Hack percentage; P is a numeric variable for protect percentage; gender is a boolean variable with Male as the baseline; education is a three-level categorical variable with HS education or less as the baseline; decisions in prior rounds are boolean with False as the baseline; and IS (internet skill), SeBIS (security behavior intention), age, time measurements, and pwd. (password) strength are numerics. Input vars shown in table for R1 2FA Decision are those that were included prior to backward selection, only S, H, P, and P:S were retained after model selection.}
\label{tab:regression:models}
\end{table}

\subsection{Results Tables}
The tables below present full regression model results for all findings presented in the text. We report the variable (one of those from Appendix~\ref{sec:appx:variables}), the log-adjusted odds ratio for the coefficient of that variable (O.R.), the 95\% confidence interval (C.I.) for that O.R., and the p-value (where the significance threshold is 0.05).

\begin{table}[h!]
\begin{tabular}{lrrr}
 \hline
 Variable & O.R. & C.I. & p-value \\ 
 \hline 
 R1 2FA Decision & 83.11 & [21.17, 326.24] & $<$0.001*\\
 \hline
\end{tabular}
\caption{Logistic regression results for 2FA behavior in RD2 modeled as a function of RD1 (R1) 2FA decision. McFadden Pseudo-$R^2$=0.353.}
\label{tab:rd2:rd1:regression}
\end{table}

\begin{table}[h!]
\begin{tabular}{lrrr}
 \hline
 Variable & O.R. & C.I. & p-value \\ 
 \hline
 R1 2FA Decision & 94.81 & [22.04, 407.8] & $<$0.001* \\ 
 R1 Login Times & 0.94 & [0.83, 0.97]  & 0.037* \\ 
 R1 Signup Time & 0.92 & [0.73, 1.16] & 0.484 \\ 
 R1 2FA:Login Times & 1.10 & [0.82, 1.47] & 0.514 \\ 
 R1 2FA:Signup Time & 2.20 & [0.32, 15.26] & 0.423 \\ 
 \hline
\end{tabular}
\caption{Logistic regression results for 2FA behavior in RD2 modeled as a function of RD1 (R1) 2FA decision and RD1 costs (mean login time and signup time in RD1, both time measures do not include time spent on 2FA screens if 2FA was enabled). McFadden Pseudo-$R^2$=0.522.}
\label{tab:rd2:threefac}
\end{table}

\begin{table}[h!]
\begin{tabular}{lrrr}
 \hline
 Variable & O.R. & C.I. & p-value \\ 
 \hline
 Endowment & 1.25 & [1.05, 1.49] & 0.015* \\ 
 IS & 1.15 & [1.03, 1.29] & 0.015* \\
 $H$ & 15.38 & [10.89, 21.73] & $<$0.001* \\ 
  \hline
\end{tabular}
\caption{Binomial logistic regression model with whether the participant made a utility-optimal decision in RD1 as the output factor and internet skill score, condition, and hack percentage as inputs, after backward selection. McFadden Pseudo-$R^2$ = 0.141.}
\label{tab:rd1:utility:regression}
\end{table}

\begin{table}[h!]
\begin{tabular}{lrrr}
 \hline
 Variable & O.R. & C.I. & p-value \\ 
 \hline
 Endowment & 0.17 & [0.01, 4.51] & 0.293 \\ 
 $H$ & 1.21 & [1.01, 1.87] & 0.041* \\ 
$P$ & 1.03 & [0.99, 1.90] & 0.109 \\ 
 SeBIS & 3.89 & [1.12, 13.46] & 0.039* \\ 
 Endowment:$H$ & 1.02 & [1.01,1.14] & 0.09 \\ 
 Endowment:$P$ & 76.49 & [0.41, 4162.32] & 0.103 \\ 
  \hline
\end{tabular}
\caption{Binomial logistic regression model with whether the participant made a utility-optimal decision in RD2 as the output factor and SeBIS score, condition, hack percentage, and protect percentage (as well as interactions between the latter three factors) as inputs, after backward selection. McFadden Pseudo-$R^2$=0.078.}
\label{tab:rd2:utility:regression}
\end{table}
\clearpage
%
\clearpage
\begin{sidewaystable}
\caption*{\large{Thought-Experiment Computations}}
\small
\centering
\begin{tabular}{cccccc}
 \hline
\backslashbox{Login Freq.}{Income (wk)}& All (\$741)  & Most (\$554) & Half (\$522) & Little (\$261) & Very Little (\$52) \\ 
 \hline
 Daily & 
 \begin{tabular}{l|l|c}
 Risk & Utility & 2FA\\
 1\% & \$0.07 & \xmark \\
 1.5\% & \$0.11 & \xmark \\
 2.7\% & \$0.19 & \xmark \\
 4.9\% & \$0.35 & \cmark \\
 20\% & \$1.42 & \cmark\\
 \end{tabular}
 &   
 \begin{tabular}{l|l|c}
 Risk & Utility & 2FA\\
 1\% & \$0.05 & \xmark \\
2.0\% & \$0.11 & \xmark \\
 3.5\% & \$0.19 & \xmark \\
6.5\% & \$0.35 & \cmark \\
 20\% & \$1.07 & \cmark\\
 \end{tabular}
 &
 \begin{tabular}{l|l|c}
 Risk & Utility & 2FA\\
 1\% & \$0.025 & \xmark \\
 2.2\% & \$0.11 & \xmark \\
 3.7\% & \$0.19 & \xmark \\
6.9\% & \$0.35 & \cmark \\
 20\% & \$1.00 & \cmark\\
 \end{tabular}
 &
 \begin{tabular}{l|l|c}
 Risk & Utility & 2FA\\
 1\% & \$0.025 & \xmark \\
  4.2\% & \$0.11 & \xmark \\
 7.4\% & \$0.19 & \xmark \\
14\% & \$0.35 & \cmark \\
 20\% & \$0.50 & \cmark\\
 \end{tabular}
 &
 \begin{tabular}{l|l|c}
 Risk & Utility & 2FA\\
 1\% & \$0.01 & \xmark \\
 20\% & \$0.10 & \xmark\\
 21\% & \$0.11 & \xmark\\
 35\% & \$0.19 & \xmark\\
 68\% & \$0.34 & \cmark\\
 \end{tabular}\\
 &&&&&\\
 $3\times$ a week & 
\begin{tabular}{l|l|c}
 Risk & Utility & 2FA\\
 1\% & \$0.07 & \xmark \\
 1.5\% & \$0.11 & \xmark \\
 2.7\% & \$0.19 & \cmark \\
 4.9\% & \$0.36 & \cmark \\
 20\% & \$1.42 & \cmark\\
 \end{tabular}
 &   
 \begin{tabular}{l|l|c}
 Risk & Utility & 2FA\\
 1\% & \$0.05 & \xmark \\
2.0\% & \$0.11 & \xmark \\
 3.5\% & \$0.19 & \cmark \\
6.5\% & \$0.35 & \cmark \\
 20\% & \$1.07 & \cmark\\
 \end{tabular}
 &
 \begin{tabular}{l|l|c}
 Risk & Utility & 2FA\\
 1\% & \$0.025 & \xmark \\
 2.2\% & \$0.11 & \xmark \\
 3.7\% & \$0.19 & \cmark \\
6.9\% & \$0.35 & \cmark \\
 20\% & \$1.00 & \cmark\\
 \end{tabular}
 &
 \begin{tabular}{l|l|c}
 Risk & Utility & 2FA\\
 1\% & \$0.025 & \xmark \\
  4.2\% & \$0.11 & \xmark \\
 7.4\% & \$0.19 & \cmark \\
14\% & \$0.35 & \cmark \\
 20\% & \$0.50 & \cmark\\
 \end{tabular}
 &
 \begin{tabular}{l|l|c}
 Risk & Utility & 2FA\\
 1\% & \$0.01 & \xmark \\
 20\% & \$0.10 & \xmark\\
 21\% & \$0.11 & \xmark\\
 35\% & \$0.19 & \cmark\\
 68\% & \$0.34 & \cmark\\
 \end{tabular}\\
 &&&&&\\
 Weekly & 
\begin{tabular}{l|l|c}
 Risk & Utility & 2FA\\
 1\% & \$0.07 & \xmark \\
 1.5\% & \$0.11 & \cmark \\
 2.7\% & \$0.19 & \cmark \\
 4.9\% & \$0.36 & \cmark \\
 20\% & \$1.42 & \cmark\\
 \end{tabular}
 &   
 \begin{tabular}{l|l|c}
 Risk & Utility & 2FA\\
 1\% & \$0.05 & \xmark \\
2.0\% & \$0.11 & \cmark \\
 3.5\% & \$0.19 & \cmark \\
6.5\% & \$0.35 & \cmark \\
 20\% & \$1.07 & \cmark\\
 \end{tabular}
 &
 \begin{tabular}{l|l|c}
 Risk & Utility & 2FA\\
 1\% & \$0.025 & \xmark \\
 2.2\% & \$0.11 & \cmark \\
 3.7\% & \$0.19 & \cmark \\
6.9\% & \$0.35 & \cmark \\
 20\% & \$1.00 & \cmark\\
 \end{tabular}
 &
 \begin{tabular}{l|l|c}
 Risk & Utility & 2FA\\
 1\% & \$0.025 & \xmark \\
  4.2\% & \$0.11 & \cmark \\
 7.4\% & \$0.19 & \cmark \\
14\% & \$0.35 & \cmark \\
 20\% & \$0.50 & \cmark\\
 \end{tabular}
 &
 \begin{tabular}{l|l|c}
 Risk & Utility & 2FA\\
 1\% & \$0.01 & \xmark \\
 20\% & \$0.10 & \xmark\\
 21\% & \$0.11 & \cmark\\
 35\% & \$0.19 & \cmark\\
 68\% & \$0.34 & \cmark\\
 \end{tabular}\\
  \hline
\end{tabular}
\caption{Utility-optimal 2FA choices for different, hypothetical, categories of MTurk users. \cmark indicates that 2FA is utility-optimal given costs from our study, while \xmark indicates that 2FA is not utility optimal. We assume that 2FA offers protection from hacking 50\% of the time.}
\label{tab:thoughtexp}
\end{sidewaystable}

\end{document}